\documentclass[a4paper]{amsart}
\usepackage[utf8]{inputenc}
\usepackage{eucal}
\usepackage{tikz-cd}
\usepackage{amssymb,amsmath,amsthm}
\usepackage{lmodern}
\usepackage[T1]{fontenc}
\usepackage[hidelinks]{hyperref}
\usepackage{mathrsfs}
\usepackage{mathtools}
\usepackage{graphicx}
\usepackage{tikz}
\usepackage{dsfont} %for the identity matrix (\mathbb{1} clashes with ams?)
\usepackage{wasysym}
\usepackage{booktabs}

\calclayout

\usepackage{thmtools}
\declaretheorem[name=Example, style=definition]{example}
\declaretheorem[name=Remark, style=definition, numbered=no]{rem}

 \newcommand{\mf}{\mathfrak} \newcommand{\mc}{\mathcal} \newcommand{\ms}{\mathsf}
\newcommand{\on}{\operatorname}

\newcommand{\la}{\langle} \newcommand{\ra}{\rangle}
\newcommand{\R}{\mathbb{R}}

\newcommand{\slot}{\;\cdot\;}

\newcommand{\w}[1]{\wedge^{\!#1}\,}
\renewcommand{\dh}{\hat d\hspace{-.5mm}}
\newcommand{\vv}[1]{\vec{#1\mkern-1mu}\mkern 1mu}

\title{Exceptional algebroids and type IIB superstrings}

\author{Mark Bugden}
\address{Current address: Department of Collective Behavior, Max Planck Institute of Animal Behavior,
Universit\"{a}tsstr. 10,
Konstanz, 78464 
Germany}
\email{mathphys@mark-bugden.com}
\author{Ond\v rej Hul\' ik}
\address{Theoretische Natuurkunde, Vrije Universiteit Brussel, Pleinlaan 2, B-1050 Brussels, Belgium\newline\indent
Institute of Particle Physics and Nuclear Physics, Faculty of Mathematics and Physics, Charles University,  V Hole\v{s}ovi\v{c}k\'{a}ch 2, 180 00 Prague 8, Czech Republic}
\email{ondra.hulik@gmail.com}
\author{Fridrich Valach}
\address{Department of Physics, Imperial College London\\Prince Consort Road, London, SW7 2AZ, UK}
\email{fridrich.valach@gmail.com}
\author{Daniel Waldram}
\address{Department of Physics, Imperial College London\\Prince Consort Road, London, SW7 2AZ, UK}
\email{d.waldram@imperial.ac.uk}

\usepackage[absolute]{textpos}
\begin{document}

\begin{textblock}{5}(14,2.5)
\noindent Imperial/TP/21/DW/3
\end{textblock}

\maketitle
\begin{abstract}
    In this note we study exceptional algebroids, focusing on their relation to type IIB superstring theory. We show that a IIB-exact exceptional algebroid (corresponding to the group $\ms E_{n(n)}\times \R^+$, for $n\le 6$) locally has a standard form given by the exceptional tangent bundle. We derive possible twists, given by a flat $\mf{gl}(2,\mathbb{R})$-connection, a covariantly closed pair of 3-forms, and a 5-form, and comment on their physical interpretation. Using this analysis we reduce the search for Leibniz parallelisable spaces, and hence maximally supersymmetric consistent truncations, to a simple algebraic problem. We show that the exceptional algebroid perspective also gives a simple description of Poisson--Lie U-duality without spectators and hence of generalised Yang--Baxter deformations.
\end{abstract}
\section{Introduction}
It has been known for some time that various classes of algebroids play an important role in string and M-theory. For instance, Courant algebroids \cite{LWX} provide many insights into the nature of string sigma models \cite{Severa1, Severa2, Severa3} and the symmetries of 10-dimensional supergravity theories \cite{HZ,CSCW0}, while also making duality symmetries (such as the Poisson--Lie T-duality of \cite{KS}) transparent. In the context of 11-dimensional supergravity and its compactifications, a different class of so-called Leibniz algebroids encodes the gauge symmetries \cite{PW,Baraglia,CSCW}.

Attempting to describe the general structure of these algebroids, and to obtain some new insights in the M-theory case, we have recently introduced \cite{one} the general notion of a $\ms G$-algebroid, depending on a choice of a specific group data set. Courant algebroids are described by taking $\ms G=\ms O(n,n)$, while using the exceptional groups (at least up to rank 6) one recovers the M-theory Leibniz algebroids. In the latter case one talks about \emph{exceptional algebroids}, or simply \emph{elgebroids}. Defining the notion of \emph{exact} elgebroids leads to two classes, related to the eleven-dimensional and type IIB supergravity.

Focusing on the M-theory case, in the paper \cite{one}, a classification result and a method for constructing Leibniz parallelisable spaces \cite{LSCW} was discussed. We also gave an algebroid definition of the general notion of Poisson--Lie U-duality, extending the construction via exceptional Drinfeld algebras introduced in \cite{Sakatani, MT} (see also \cite{BTZ,Sakatani2} for discussion in the context of type IIB). The aim of the present note is to develop the corresponding theory for the type IIB case.

Specifically, we first introduce IIB-exact elgebroids and study their local classification, including the possible twists, their Bianchi identities and physical interpretation in supergravity. We then proceed to the construction of a class of IIB-exact elgebroids as the pull-back of some simple algebraic data. These are in one-to-one correspondence with exceptional Leibniz parallelisations and hence define maximally supersymmetric consistent truncations, where the underlying algebra encodes the \emph{embedding tensor} of the corresponding gauged supergravity. We prove a structure theorem for such constructions, giving a new perspective on and slightly refining a result obtained by Inverso \cite{Inverso} (see also \cite{BHL} for the $n=4$ case). We explain how several standard examples fit into the formalism, and note how Poisson--Lie U-duality and generalised Yang--Baxter deformations can be simply described in this language. We conclude with a brief summary of the results of this note and \cite{one}.

As might be expected, the type-IIB story is slightly more technically involved than the M-theory case. It also occasionally leads to small surprises, such as the possibility of twists of the bracket by a pair of vector fields -- which, although ultimately disappearing due to the Jacobi identity, enters in the analysis of Section \ref{sec:leib} (and in particular leads to a certain unimodularity-type condition when considering Leibniz parallelisations). One also notes the natural emergence of a flat $\mf{gl}(2,\R)$-connection which is linked to the axion, dilaton, and the warp factor. 
% (The apparent discrepancy between the numbers of degrees of freedom is explained at the end of Section \ref{sec:elg}.)

\subsection*{Acknowledgements}
    F.V.\ was supported by the Early Postdoc Mobility grant P2GEP2\underline{\phantom{k}}188247 of the Swiss National Science Foundation. D.W.\ was supported in part by the STFC Consolidated Grant ST/T000791/1 and the EPSRC New Horizons Grant EP/V049089/1. 

\section{Algebraic prelude}
   By way of prelude, we recall the algebraic data one needs to define an elgebroid following \cite{one}. Let $n\in\{2,\dots,6\}$. We then take the group $\ms E_{n(n)}$, together with a pair of its representations $E$ and $N$, from the following table. \vspace{.2cm}
    \begin{center}\begin{tabular}{@{}lccccc@{}}
        \toprule
         $n$ & 2 & 3 & 4 & 5 & 6\\
         \midrule
        $\ms E_{n(n)}$ & $\ms{SL}(2,\R)\times\R^+$ & $\ms{SL}(3,\R)\times\ms{SL}(2,\R)$&$\ms{SL}(5,\R)$&$\ms{Spin}(5,5)$ & $E_{6(6)}$\\
        $E$ & $\mathbf{1}_{\mathbf{1}}\oplus\mathbf{2}_{-\mathbf{1}}$ & $(\mathbf{3},\mathbf{2})$ & $\mathbf{10}$ & $\mathbf{16}$ & $\mathbf{27}$\\
        $N$ & $\mathbf{1}_{\mathbf{0}}$ & $(\mathbf{3}',\mathbf{1})$ & $\mathbf{5}'$ & $\mathbf{10}$ & $\mathbf{27}'$ \\
%        $\lambda$ & $2/3$ & $1/2$ & $2/5$ & $5/16$ & $2/9$
        \bottomrule
    \end{tabular}\end{center}
   \vspace{.2cm}
    These groups (apart from $\R$) can be seen as split real forms of complex semisimple Lie algebras. We will be interested in the group $\ms G:=\ms E_{n(n)}\times\R^+$, where the extra $\R^+$ factor acts on $E$ and $N$ with weights 1 and 2, respectively. Note that the two representations $E$ and $N$ are part of the general sequence that appears in the tensor heirarchy \cite{dWNS}. More details concerning these groups and representations can be found in the Appendix.

    Importantly, $N$ can be seen as a subrepresentation of the second symmetric power of $E$. Taking suitable multiples of this embedding and of the corresponding projection, we obtain two $\ms G$-equivariant maps $N\to E\otimes E$ and $E\otimes E\to N$, satisfying the following property. Define $\pi'\colon \on{End}(E)\to \on{End}(E)$ as the partial dual of the composition $E\otimes E\to N\to E\otimes E$,\footnote{i.e., denoting the map $E\otimes E\to E\otimes E$ by $\mu$, we set $\la \pi'(\xi,u), v\otimes \eta\ra:=\la \mu(u\otimes v),\xi\otimes\eta\ra$ for $u,v\in E$, $\xi,\eta\in E^*$ (we use $\on{End}(E)\cong E^*\otimes E$)} and set $\pi:=1-\pi'$. We then have 
    \begin{equation}\label{eq:meta}
        \on{Im}(\pi)\subset \mf g\subset \on{End}(E),
    \end{equation}
    where $\mf g$ is the Lie algebra of $\ms G$.
    
    To simplify the notation, we shall use ``target subscripts'' when referring to the two maps $E\otimes E\to N$, $N\to E\otimes E$, or their (partial) duals  (e.g.\ $(u\otimes v)_N$ for the image of $u\otimes v$ under the former map or $(\xi\otimes n)_E$ for $E^*\otimes N\to E$, a partial dual of the latter map). 
    
    Using these maps we can also define the notions of Lagrangian and co-Lagrangian subspaces. Namely, a subspace $V\subset E$ is called \emph{Lagrangian} if $(V\otimes V)_N=0$ and if it cannot be further enlarged, preserving this property. Similarly, $V\subset E$ is \emph{co-Lagrangian} if $(V^\circ\otimes V^\circ)_{N^*}=0$ and if $V$ has no proper subspace with the same property. (Here $V^\circ\subset E^*$ is the annihilator of $V$.)
    
    We note a few important and useful observations:
    \begin{itemize}
        \item There are precisely two classes of co-Lagrangian subspaces -- those of codimension $n$, called \emph{type M}, and those of codimension $n-1$, called \emph{type IIB}.
        \item $V$ is co-Lagrangian if and only if $(V^\circ\otimes N)_E= V$.
        \item In the case $n>2$, for any $A\in\on{End}(E)$ we have $\on{Tr}_E\pi(A)=\lambda\on{Tr}_EA$, with $\lambda=-\tfrac{\dim E}{9-n}$.
    \end{itemize}
\section{Elgebroids}\label{sec:elg}
    We now start by defining the central object of our study, introduced in \cite{one} under the name \emph{exceptional algebroid} or \emph{elgebroid}. In order to keep the notation simple, we shall (by a mild abuse of notation) use the same letters to denote representations $E$, $N$, and the corresponding associated bundles.
    
    An \emph{elgebroid} is given by a principal $\ms G$-bundle over a manifold $M$, together with a set of structures on the associated bundles $E\to M$ and $N\to M$, namely
    \begin{itemize}
        \item an ($\R$-bilinear) bracket $[\cdot,\cdot]\colon \Gamma(E)\otimes \Gamma(E)\to\Gamma(E)$,
        \item a vector bundle map $\rho\colon E\to TM$, called the \emph{anchor},
        \item a $\R$-linear operator $\mc D\colon \Gamma(N)\to\Gamma(E)$.
    \end{itemize}
    This data is subject to some conditions:
    \begin{itemize}
        \item $E$ is a Leibniz algebroid, i.e.\ for all $u,v,w\in\Gamma(E)$ and $f\in C^\infty(M)$ we have
        \begin{equation}\label{eq:leibniz}
            [u,[v,w]]=[[u,v],w]+[v,[u,w]],\qquad [u,fv]=f[u,v]+(\rho(u)f)v.
        \end{equation}
        \item Define $\dh f:=\rho^t(df)\in \Gamma(E^*)$, where $\rho^t\colon T^*M\to E^*$ is the transpose (dual map) of $\rho$. We require that the symmetric part of the bracket is governed by the conditions (here $n\in \Gamma(N)$)
        \begin{align}
            [u,v]+[v,u]=\mc D(u\otimes v)_N,\qquad \mc D(fn)=f\mc Dn+(\dh f\otimes n)_E.
        \end{align}
        \item The bracket preserves the $\ms G$-structure.
    \end{itemize}
    
    Note that the map $\mc D$ is fully determined in terms of the bracket -- nevertheless, it is still convenient to keep it as part of the definition.
    
    As a simple consequence of the axioms, one has
    \begin{equation}\label{eq:cons}
        \rho([u,v])=[\rho(u),\rho(v)]\qquad [fu,v]=f[u,v]-\pi(\dh f\otimes u)v.
    \end{equation}
    For instance, the first equation can be obtained by using $[x,fy]=f[x,y]+(\rho(x)f)y$ on both sides of $[u,[v,fw]]=[[u,v],fw]+[v,[u,fw]]$.
    
    Consequently, $\rho\circ \mc D=0$ and thus also $\rho(\dh f\otimes n)_E=0$. Using the fact that $(\on{Ker}\rho)^\circ=\on{Im}\rho^t$ we then get the chain complex (using the map $N\to E\otimes E$)
    \begin{equation}\label{eq:complex}
        T^*M\otimes N\xrightarrow{\rho^t} E\xrightarrow{\rho} TM\to 0.
    \end{equation}
    If this is an exact sequence, we say the elgebroid is \emph{exact}. Note that in particular this implies $\on{Ker}\rho$ is co-Lagrangian \cite{one}. Thus, we again distinguish \emph{M-exact} and \emph{IIB-exact} elgebroids, depending on whether $\dim M=n$ or $\dim M=n-1$, respectively.
    
    It was shown in \cite{one} that any M-exact elgebroid is locally of the standard form
    \[E\cong TM\oplus \w{2}T^*M\oplus\w{5}T^*M,\]
    with the anchor given by the projection onto the first factor and the bracket being
    \[[X+\sigma_2+\sigma_5,X'+\sigma_2'+\sigma_5']=\mc L_X X'+(\mc L_X \sigma_2'-i_{X'}d\sigma_2)+(\mc L_X \sigma_5'-i_{X'}d\sigma_5-\sigma_2'\wedge d\sigma_2).\]
\section{Exceptional tangent bundle}\label{sec:etb}
    We shall now prove the following statement\vspace{.2cm}

    \emph{Every IIB-exact elgebroid is locally of the form of the \emph{(type IIB) exceptional tangent bundle} \cite{Hull, CSCW, BCKT, CSCW2, BMP,HS2}, i.e.
    \begin{equation}\label{eq:bundle}
        E\cong TM\oplus (S\otimes T^*M)\oplus \w{3}T^*M\oplus (S\otimes \w{5}T^*M),
    \end{equation}
    where $S:=\R^2$, the anchor is given by the projection onto $TM$, (the $\ms G$-structure is the one described in the Appendix) and the bracket is
    \begin{align}
        [X+\vv\sigma_1+\sigma_3+\vv\sigma_5,X'+\vv\sigma_1'+\sigma_3'+\vv\sigma_5']&=\mc L_X X'+(\mc L_X\vv \sigma_1'-\iota_{X'}d\vv\sigma_1)+(\mc L_X \sigma_3'-\iota_{X'}d\sigma_3+\epsilon_{ij}d\sigma_1^i\wedge {\sigma'_1}^{\!j})\nonumber\\
        &+(\mc L_X\vv\sigma_5'-\iota_{X'}d\vv\sigma_5+d\sigma_3\wedge \vv\sigma_1'-d\vv\sigma_1\wedge\sigma_3').\label{eq:bracket}
    \end{align}}
    
    It is convenient to begin the proof by relaxing the first requirement in \eqref{eq:leibniz} and instead adopting the first equation in \eqref{eq:cons} -- we obtain a ``weaker'' structure called a \emph{pre-elgebroid} in \cite{one}. We will first constrain the form of this object, and later on apply the first condition in \eqref{eq:leibniz}, i.e.\ the \emph{Jacobi identity}, to give a true elgebroid. 
    
    Supposing $E$ is a IIB-exact\footnote{IIB-exactness is defined in the same way as for elgebroids.} pre-elgebroid, we note that locally there exists a vector bundle isomorphism \eqref{eq:bundle}, which preserves the anchors and the $\ms G$-structure. This follows from the facts that both $(S\otimes T^*M)\oplus \w{3}T^*M\oplus (S\otimes \w{5}T^*M)$ and $\on{Ker}\rho$ are type IIB co-Lagrangian and that all type IIB co-Lagrangian subspaces are related by a $\ms G$-transformation (c.f.\ \cite{one}).
    
    Let us therefore make this identification, implying in particular that the maps $E\otimes E\to N$, $N\to E\otimes E$, as well as the other bundles ($N$, the adjoint, etc.) take the form in the Appendix. It remains to restrict the form of the bracket.
    
    Choosing coordinates on $M$, locally we get a trivialisation $E \cong [T\oplus (S\otimes T^*)\oplus \w{3}T^*\oplus (S\otimes \w{5}T^*)]\times M$, with $T:=\R^{n-1}$. In particular, sections of $E$ (and similarly for the other bundles) can be seen as functions on $M$ valued in the vector space $T\oplus (S\otimes T^*)\oplus \w{3}T^*\oplus (S\otimes \w{5}T^*)$. Crucially, it follows from the definition of elgebroid that the expression $[u,v]-\rho(u)v+\pi(\hat d u)v$ is tensorial in $u$ and $v$ (here $\rho(u)$ and $d$ act only on the $C^\infty(M)$-part of the sections, leaving the $T\oplus (S\otimes T^*)\oplus \w{3}T^*\oplus (S\otimes \w{5}T^*)$-part intact). Since the bracket preserves the $\ms G$-structure, we get that it can be written as
    \begin{equation}\label{eq:trivialis}
        [u,v]=\rho(u)v-\pi(\hat d u)v+A(u)\cdot v,
    \end{equation}
    where $A$ is, at each point on $M$, a map \[T\oplus (S\otimes T^*)\oplus \w{3}T^*\oplus (S\otimes \w{5}T^*) \to \R\oplus \mf{gl}(T)\oplus \mf{sl}(S)\oplus (S\otimes \w{2}T)\oplus (S\otimes \w{2}T^*)\oplus \w{4}T\oplus\w{4}T^*.\]
    Here and henceforth, $\cdot$ will denote the action of a Lie algebra (or group) on a given module.

    Writing $u=X+\sigma$, with $\sigma=\vv\sigma_1+\sigma_3+\vv\sigma_5$ (the arrow signifying that the tensor is valued in $S$), the first two terms in \eqref{eq:trivialis} can be written as $\rho(u)v-\pi(\hat d u)v=\mc L_Xv-(d\sigma)\cdot v$, and thus correspond precisely to the bracket \eqref{eq:bracket}. The strategy is now to constrain the form of the tensor $A$ and then show how to use the freedom in the identification \eqref{eq:bundle} to locally gauge $A$ to zero. In the course of this process, we will naturally derive the possible twists and their Bianchi identities. Note that this step is analogous to the analysis of general ``deformations'' in \cite{CGI,CdFPSW,Inverso}. 
    
    First, we use the fact that for $u,v\in\Gamma(E)$ constant we have $0=[\rho(u),\rho(v)]=\rho([u,v])=\rho(A(u)\cdot v)$ to get $\on{Im}A\subset \mf n=\mf{gl}(S)\oplus (S\otimes \w{2}T^*)\oplus\w{4}T^*$ (see the Appendix).
    
    Second, let us write $\mc Dn=(\hat d n)_E+B(n)$, for $B\colon N\to E$. A straightforward but slightly tedious calculation reveals that the condition $B(u\otimes v)_N=A(u)\cdot v+A(v)\cdot u$ restricts $A$ to have the form
    \begin{align*}
             A(u)=(\iota_X\mc F_1+\epsilon_{jk}\iota_{\psi^i}\sigma_1^k \, e^j_i)+(\iota_X\vv F_{\!3}-\mc F_1\wedge \vv\sigma_1+\iota_{\vv\psi}\sigma_3)+(\iota_X F_5+\epsilon_{ij} F_{\!3}^i\wedge\sigma_1^j-\on{Tr}\mc F_1\wedge\sigma_3+\epsilon_{ij}\iota_{\psi^i}\sigma_5^j),
    \end{align*}
    where $e_i^j$ is the basis of $\mf{gl}(S)$ and we have the twists given by
    \[\mc F_1\in \mf{gl}(S)\otimes T^*, \quad\vv F_{\!3}\in S\otimes\w{3}T^*,\quad F_5\in\w{5}T^*,\quad \vv\psi \in S\otimes T.\]
    Most of these twists have direct physical interpretations, see the discussion below. A notable exception is given by the pair of vectors $\vv\psi$, which is non-physical and will disappear upon imposing the Jacobi identity.\footnote{An easy way to see the appearance of these vectors in the calculation is to look at the case $n=2$, where we have $A\colon T\oplus (S\otimes T^*)\cong \R\oplus S\to \mf{gl}(S)$. The only restriction on $A$ set by $B(u\otimes v)_N=A(u)\cdot v+A(v)\cdot u$ is that $A|_S\colon S\to\mf{gl}(S)$ defines a skew symmetric bracket on $S$, i.e.\ $S$ is a 2-dimensional Lie algebra. The vectors $\vv\psi$ then parametrise the Lie bracket. The remaining part $A|_\R\colon \R\to \mf{gl}(S)$ is unrestricted and corresponds to $\mc F_1$.} It is intriguing that they are allowed however by the weaker pre-elgebroid structure. 
    
    Indeed, consider now the condition $[\mc Dn,w]=0$, which follows from the Jacobi identity by setting $u=v$. Taking $w=\vv\sigma_1\in\Gamma(S\otimes T^*M)$ and $n=\vv n_0\in\Gamma(S \times M)$, the $S\otimes T^*M$-part of $[\mc D \vv n_0,\vv\sigma_1]$ is $\epsilon_{ij}[\iota_{\vv\psi}(dn^j_0+\mc F_1 n^j_0)]\sigma_1^i$. Thus, its vanishing requires $\vv\psi=0$.
    
    A simple calculation then reveals
    \[[\mc D \vv n_0,\vv\sigma_1]=-\epsilon_{ij}\sigma_1^i\wedge(d\mc F_1+\tfrac12 [\mc F_1,\mc F_1])^j_kn_0^k-\vv\sigma_1\wedge \epsilon_{ij}(d\vv F_{\!3}+\mc F_1\wedge \vv F_{\!3})^in_0^j.\]
    Thus, Jacobi identity implies that $\mc F_1$ is a flat $\mf{gl}(S)$-connection and $\vv F_{\!3}$, seen as living in the vector representation of $\mf{gl}(S)$, is covariantly closed.\footnote{The remaining flux $F_5$ is automatically covariantly closed, since $\dim M\le 5$. We will see below that the Jacobi identity doesn't impose any further constraints on the fluxes, apart from those just mentioned.} These are the Bianchi identities.
    
    As the next step, note that our above identification \eqref{eq:bundle} was not unique. Two such identifications are related by a $\ms G$-transformation which preserves the anchor, i.e.\ by $g\in \ms N$, where $\ms N$ corresponds to the Lie algebra $\mf n$. Let us use the notation
    \begin{equation}\label{eq:bkt}
        [X+\sigma,\;\cdot\;]_{\mc F_1,\vv F_{\!3},F_5}:=\mc L_X+(-d\sigma+\iota_X\mc F_1+(\iota_X \vv F_3-\mc F_1\wedge \vv\sigma_1)+(\iota_X F_5+\epsilon_{ij}F_3^i\wedge \sigma_1^j-\on{Tr}\mc F_1\wedge\sigma_3))\cdot
    \end{equation}
    Under a local $\mf n$-transformation $\varphi$ the bracket changes by
    \[\delta[\slot,\slot]_{\mc F_1,\vv F_{\!3},F_5}:=\varphi[\slot,\slot]_{\mc F_1,\vv F_{\!3},F_5}-[\varphi\slot,\slot]_{\mc F_1,\vv F_{\!3},F_5}-[\slot,\varphi\slot]_{\mc F_1,\vv F_{\!3},F_5}=[\slot,\slot]_{\mc F_1+\delta\mc F_1,\vv F_{\!3}+\delta \vv F_{\!3},F_5+\delta F_5}.\]
    For $\varphi=c\in \Gamma(\mf{gl}(S)\times M)$, this gives
    \begin{align}\label{eq:gauge1}
        \delta \mc F_1=[c,\mc F_1]-dc\qquad \delta \vv F_{\!3}=c \, \vv F_{\!3},\qquad \delta F_5=(\on{Tr}c) F_5,
    \end{align}
    i.e., $\mc F_1$ transforms as a connection while $\vv F_{\!3}$ and $F_5$ live in the vector and trace representations of $\mf{gl}(S)$, respectively. For $\varphi=\vv A_2\in\Gamma(S\otimes \w{2}T^*M)$ we have
    \begin{align}
        \delta \mc F_1=0,\qquad \delta \vv F_{\!3}=-d\vv A_2-\mc F_1\wedge \vv A_2,\qquad \delta F_5=\epsilon_{ij}F_{\!3}^i\wedge A_2^j,
    \end{align}
    while for $\varphi=A_4\in\Gamma(\w{4}T^*M)$ we get
    \begin{align}\label{eq:gauge3}
        \delta \mc F_1=0,\qquad \delta \vv F_{\!3}=0,\qquad \delta F_5=-dA_4-(\on{Tr}\mc F_1)\wedge A_4.
    \end{align}
    
    In particular, we can always (locally) gauge the fluxes away, obtaining $A=0$, which concludes the proof. Conversely, it follows that the twisted bracket satisfies the Jacobi identity for any $\mc F_1$, $\vv F_{\!3}$, $F_5$ satisfying the Bianchi identities (since the untwisted bracket does).
    
    To get a full global picture for exact elgebroids, one should study the patching of these local descriptions together. We leave this to a later work.
    
    Finally, notice that for any IIB-exact elgebroid and any section $u\in\Gamma(\on{Ker}\rho)$ the operator $[u,\slot]$, which can be seen as a section of $E^*\otimes E$, is traceless -- we will use this fact in Section \ref{sec:leib}.
    \begin{rem}
        Let us now comment on the physical interpretation of the twists $\mc F_1$, $\vv F_{\!3}$, $F_5$.
        
        First, recall that IIB-exact elgebroids encode the symmetries of the restriction of  10-dimensional type IIB supergravity to a $n-1$-dimensional space, with the warp factor included. The bosonic field content of the restricted supergravity consists of 
        \begin{itemize}
            \item three scalars: the warp factor, the axion and the dilaton; the last two can be seen as parametrising the $\ms{SL}(2,\R)/\ms U(1)$ coset space and thus carrying a nonlinear action of the S-duality group $\ms{SL}(2,\R)$; the warp factor is a scalar w.r.t.\ this group
            \item two 2-forms, forming a doublet under the S-duality group
            \item one 4-form, corresponding to a singlet of the S-duality group.
        \end{itemize}
        Naively, one might expect the twists are the ``field strengths'' of these fields (in particular the derivative of the warp factor should correspond to the $\R'$-part of $\mc F_1$) -- it is thus at first blush surprising to see that we have only the axion and the dilaton to account for the 3-dimensional Lie algebra $\mf{sl}(2,\R)$. Let us briefly explain the apparent discrepancy, by looking in a little more detail into the correspondence between twists and fields. 
        
        As was derived above, a IIB-exact elgebroid twisted by $\mc F_1$, $\vv F_{\!3}$, $F_5$ is equivalent to the one twisted by $\mc F_1'$, $\vv F_{\!3}'$, $F_5'$, provided the two sets of twists are related by a local $\ms N$-transformation.\footnote{We are ignoring the global issues here, focusing instead on a small patch of the base manifold.}
        However, suppose we equip such elgebroid with a field content of the lower-dimensional supergravity. This is equivalent to specifying a generalised metric on $E$ and crucially determines a specific choice of twists -- for instance, denoting the doublet of 2-forms by $\vv B_2$, we see that the combination $\vv F_{\!3}+d\vv B_2$ is invariant under the $\ms N$-action. There is an important exception, though, given by the fact that the $\ms U(1)\subset \ms{SL}(2,\R)$ subgroup leaves the generalised metric invariant, while it does change the fluxes. A choice of generalised metric on a IIB-exact elgebroid thus determines the fluxes, but only up to a local $\ms U(1)$-transformation. This explains the above discrepancy between the number of fields and fluxes, and reflects the coset structure for the axion-dilaton space. Finally note that in most physics applications one assumes the warp factor is globally defined and so the $\R'$-part of $\mc F_1$ can be globally gauged away to zero using the $\ms N$-action. 
    \end{rem}
    
    For completeness and future reference, let us write here the full form of the twisted bracket \eqref{eq:bkt}:
    \begin{align*}
        [X+\vv\sigma_1+\sigma_3+&\vv\sigma_5,X'+\vv\sigma_1'+\sigma_3'+\vv\sigma_5']_{\mc F_1,\vv F_{\!3},F_5}=\mc L_X X'+[\mc L_X\vv\sigma_1'-\iota_{X'}d\vv\sigma_1+(\iota_X\mc F_1)\vv\sigma_1'-\iota_{X'}(\mc F_1\wedge \vv\sigma_1)\\
        &+\iota_{X'}\iota_X\vv F_{\!3}]+[\mc L_X\sigma_3'-\iota_{X'}d\sigma_3-\epsilon_{ij}(\sigma'_1)^i\wedge d\sigma_1^j-\epsilon_{ij}(\sigma_1')^i\wedge (\mc F_1\wedge \vv\sigma_1)^j+(\iota_X\on{Tr}\mc F_1)\sigma_3'\\
        &-\iota_{X'}(\on{Tr}\mc F_1\wedge\sigma_3)+\epsilon_{ij}(\sigma'_1)^i\wedge\iota_X F_3^j+\epsilon_{ij}\iota_{X'}(F_3^i\wedge \sigma_1^j)+\iota_{X'}\iota_X F_5]+[\mc L_X\vv\sigma_5'+\vv\sigma_1'\wedge d\sigma_3\\
        &-\sigma_3'\wedge d\vv\sigma_1-\sigma_3'\wedge\mc F_1\wedge\vv\sigma_1+\vv\sigma_1'\wedge\on{Tr}\mc F_1\wedge\sigma_3+(\iota_X\mc F_1)\vv\sigma_5'+(\iota_X\on{Tr}\mc F_1)\vv\sigma_5'+\sigma_3'\wedge \iota_X \vv F_{\!3}\\
        &-\vv\sigma_1'\wedge\epsilon_{ij}F_3^i\wedge \sigma_1^j-\vv\sigma_1'\wedge\iota_X F_5],
    \end{align*}
    where $\mc F_1\in\Omega^1(M)\otimes\mf{gl}(S)$, $\vv F_{\!3}\in\Omega^3(M)\otimes S$, $F_5\in\Omega^5(M)$ satisfy the Bianchi identities
    \[d\mc F_1+\tfrac12 [\mc F_1,\mc F_1]=0,\qquad d\vv F_{\!3}+\mc F_1\wedge \vv F_{\!3}=0\]
    and their gauge transformations are given by formulas \eqref{eq:gauge1} -- \eqref{eq:gauge3}. (Recall also that $\dim M\le 5$.)
\section{Elgebras}
    In the previous Section we dealt with elgebroids which are directly linked to the physics of type IIB compactifications. We now examine a very different class of elgebroids, namely those for which the base manifold $M$ is a point -- such structures are called \emph{elgebras}.
    
    In other words, an elgebra is equivalent to a bilinear bracket on the representation space $E$ of the group $\ms G=\ms E_{n(n)}\times \R^+$, satisfying
    \begin{itemize}
        \item $[u,[v,w]]=[[u,v],w]+[v,[u,w]]$
        \item $[u,v]+[v,u]=\mc D(u\otimes v)_N$ for some linear map $\mc D\colon N\to E$
        \item $[u,\slot]\in \mf{g}\subset \on{End}E$.
    \end{itemize}
    Note that an elgebra (and more generally any ${\ms G}$-algebra \cite{one}) is an example of a ``symmetric enhanced Leibniz algebra'' as introduced by Strobl and Wagemann \cite{Strobl,SW}. Physically, the structure constants of an elgebra specify an embedding tensor describing the gauging of a maximally supersymmetric supergravity theory in $11-n$ spacetime dimensions \cite{Samtleben}. 
    
    %From this it follows that for any $u\in E$, $n\in N$, 
    %\[[\mc D n,u]=0,\qquad [u,\mc D n]=\mc D[u,n],\]
    %where in the latter expression we use the fact that $[u,\slot]\in\mf g$ acts on any representation space of $G$. Consequently, 
    It follows from the definition that $\mf g_E:=E/\on{Im}\mc D$ is a Lie algebra. Similarly, any subelgebra $V\subset E$ gives rise to a Lie subalgebra $\mf g_V:=V/\on{Im}\mc D\subset \mf g_E$. Conversely, any Lie subalgebra of $\mf g_E$ lifts to a unique subelgebra $V\subset E$ containing $\on{Im}\mc D$. We shall denote by $\ms G_E$ the 1-connected Lie group with Lie algebra $\mf g_E$. Note that this group acts on $E$, preserving the bracket.

\section{Leibniz parallelisations}\label{sec:leib}
    We now focus on the construction of a class of IIB-exact elgebroids, starting from an algebraic data including an elgebra $E$. First, note that any elgebroid $E'\to M'$ is uniquely determined by specifying the bundles $E'\to M'$, $N'\to M'$ (and the $\ms G$-structure), the anchor, and the bracket of constant sections, w.r.t.\ some trivialisation of $E'$. 
    
    Starting with an elgebra $E$ and a compact manifold $M'$, we construct the product bundles $E':=E\times M'$, $N':=N\times M'$. We can then investigate whether there exists an anchor map on $E'$ such that there is an elgebroid structure on $E'$ whose bracket on constant sections reproduces the one on $E$. Such an elgebroid is called \emph{Leibniz parallelisable}. As shown in \cite{LSCW}, these spaces correspond to consistent truncations to a $(11-n)$-dimensional maximally supersymmetric theory with an embedding tensor defined by the elgebra $E$.  

    Since the anchor on $E\times M'$ corresponds to an action of $E$ on $M'$ (i.e., a bracket preserving map $E\to \mf X(M')$), we are lead to the following question:\vspace{.2cm}
    
    \emph{When does an action of an elgebra on a compact (connected) manifold define a IIB-exact elgebroid?}\vspace{.2cm}
    
    Let us call the action $\chi\colon E\to \mf{X}(M')$. Exactness in particular implies that this is a transitive action of the elgebra $E$ ($\chi$ is surjective at every point). Since $\chi([u,v]+[v,u])=0$, we get $\chi(\on{Im}\mc D)=0$ and thus we have an induced transitive action of $\mf g_E$ on $M'$. Therefore, since $M'$ is compact, there exists a Lie subalgebra $\tilde{\mf g}\subset \mf g_E$ such that $M'\cong \ms G_E/\widetilde{\ms G}$, where $\widetilde{\ms G}\subset \ms G_E$ is a subgroup corresponding to $\tilde{\mf g}$. Lifting $\tilde{\mf g}$ to $E$, we get a corresponding subelgebra $V\subset E$, containing $\on{Im}\mc D$. Consequently, any Leibniz parallelisable space corresponds to some pair $V\subset E$ of an elgebra and its subelgebra (containing $\on{Im}\mc D$).\footnote{Note that, for a given $E,V$, we can have more Leibniz parallelisable spaces because there is often some (small) freedom in choosing the subgroup $\ms G_V\subset \ms G_E$ corresponding to the Lie subalgebra $\mf g_V$. For instance, if $\mf g_V=0$, we can take $\ms G_V$ to be any discrete subgroup of $\ms G_E$.} We now prove the following claim. Note that a closely analogous result was derived (using different methods) in \cite{Inverso}.\vspace{.2cm}
    
    \emph{Let $n>2$. Suppose $E$ is an elgebra and $V\subset E$ is a subelgebra containing $\on{Im}\mc D$, and for which there exists a closed subgroup $\ms G_V\subset \ms G_E$. Then this defines a Leibniz parallelisable IIB-exact elgebroid iff $V$ is type IIB co-Lagrangian and $\on{Tr}_E\on{ad}_v=\tfrac{\lambda}{\lambda-1}\on{Tr}_V\on{ad}_v$ for all $v\in V$.}\vspace{.2cm}
    
    First, IIB-exactness translates into the transitivity of $\chi\colon E\to\mf{X}(M')$ and the fact that $\on{Ker}\chi$ is type IIB co-Lagrangian at every point on $M'\cong \ms G_E/\ms G_V$. Since $\on{Ker}\chi$ is related to $V$ by a $\ms G_E$-transformation, we get that this is equivalent to $V$ being type IIB co-Lagrangian. Assuming this, it remains to check that the induced bracket on $E'=E\times M'$, which necessarily takes the form
    \begin{equation}\label{eq:pulledbackbracket}
        [u,v]'=[u,v]+\chi(u)v-\pi(\hat d u)v, \qquad u,v\in \Gamma(E')\cong C^\infty(M')\otimes E,
    \end{equation}
    together with $\mc D'n=\mc Dn+(\hat d n)_E$, satisfy the axioms of an elgebroid.\footnote{Note that here we have identified $\rho'$ with $\chi$.}
    This is immediate for all the conditions except for the Jacobi identity. To deal with the latter, we proceed as follows.
    
    A simple calculation shows that, taking general sections $u,v\in \Gamma(E')$, we have 
    \[[\rho'(u),\rho'(v)]'-\rho'([u,v]')=-\chi(((\hat d u\otimes v)_N)_E),\] 
    which vanishes since $\on{Ker}\chi$ is co-Lagrangian. We thus get a (IIB-exact) pre-elgebroid structure. As was shown above, this in turn (locally) fixes the form of the bracket, up to a possible twist by $\mc F_1$, $\vv F_{\!3}$, $F_5$, and the pair of vectors $\vv\psi$.\footnote{Note that we are not yet allowed to assume that these (form) twists satisfy the Bianchi identities. As was shown in Section \ref{sec:etb}, these conditions will follow automatically once we prove the Jacobi identity.}
    
    As discussed at the end of Section \ref{sec:etb}, in order to get a proper elgebroid we need to have that 
    \begin{equation}\label{eq:trace}
        \on{Tr}_E([u,\slot]')=0,\qquad \forall u\in\Gamma(\on{Ker}\rho').
    \end{equation} Assuming for the moment that this is the case, we get $\vv\psi=0$ as a consequence. In other words, locally the bracket can be put in the form \eqref{eq:bkt}, for some $\mc F_1$, $\vv F_{\!3}$, $F_5$. Crucially, a direct calculation shows that for this bracket, the \emph{Jacobiator}
    \[J(u,v,w):=[u,[v,w]]-[[u,v],w]-[v,[u,w]], \qquad u,v,w\in\Gamma(E')\]
    is a tensor. Since $E$ is an elgebra, $J$ vanishes on constant sections, and so the tensoriality of $J$ implies that the Jacobi identity on $E'$ holds identically.
    
    To finish the proof, we need to check when \eqref{eq:trace} holds. For $u\in \Gamma(\on{Ker}\rho')$ equation \eqref{eq:pulledbackbracket} implies $[u,\slot]'=[u,\slot]-\pi(\hat d u)$. Using $\on{Tr}_E\circ\,\pi=\lambda\on{Tr}_E$, we get the condition
    \begin{equation}\label{eq:aux}
        \on{Tr}_E [u_{[g]},\cdot]=\lambda\on{Tr}_E(\hat d u)_{[g]},\qquad \forall u\in\Gamma(\on{Ker}\rho'),\; g\in \ms G_E.
    \end{equation}
    Here $u_{[g]}$ is the value of $u$ at the point in $\ms G_E/\ms G_V$ given by the element $g$.
    
    Let us now simplify the RHS, writing $\on{Tr}_E(\hat d u)_{[g]}=\on{Tr}_E(x\mapsto \rho'_{[g]}(x)u)$.
    If $x\in\on{Ker}\rho'_{[g]}=g\cdot V$, then $\rho'_{[g]}(x)u$ vanishes.\footnote{Recall that $\cdot$ denotes the Lie group action (in this case of $\ms G_E$ on $E$). If $\mc D=0$ and thus $E=\mf g_E$, this action coincides with the usual adjoint action.} Otherwise we write $\rho'_{[g]}(x)u=\tfrac{d}{dt}\big|_{t=0}u_{[e^{-t\tilde x}g]}$, where $\tilde x$ is the image of $x\in E$ in $\mf g_E=E/\on{Im}\mc D$. Since $\rho'(u)=0$, we have $u_{[e^{-t\tilde x}g]}=e^{-t\tilde x}g\cdot \gamma(t)$, with $\gamma(t)\in V$ for all $t$. For $x\notin\on{Ker}\rho'_{[g]}$ we thus get \[\rho'_{[g]}(x)u=\tfrac{d}{dt}\big|_{t=0}e^{-t\tilde x}g\cdot \gamma(t)=-[x, u_{[g]}]+g\cdot \gamma'(0).\]
    Since the last term lies in $\on{Ker}\rho'_{[g]}$, we can write $\on{Tr}_E(\hat d u)_{[g]}=-\on{Tr}_{E/\on{Ker}\rho'_{[g]}}(x\mapsto [x,u_{[g]}])$. As the action of $\ms G_E$ preserves the bracket on $E$, we see that \eqref{eq:aux} is equivalent to $\on{Tr}_E [v,\slot]=-\lambda \on{Tr}_{E/V}[\slot,v]$, $\forall v\in V$. Note that $\on{Im}\mc D\subset V$ implies $\on{Tr}_{E/V}[\slot,v]=-\on{Tr}_{E/V}[v,\slot]$. Using $\on{Tr}_{E/V}\on{ad}_v=\on{Tr}_E\on{ad}_v-\on{Tr}_V\on{ad}_v$, we thus finally obtain the condition $\on{Tr}_E\on{ad}_v=\tfrac{\lambda}{\lambda-1}\on{Tr}_V\on{ad}_v$, $\forall v\in V$, finishing the proof.

    The above statement reduces the search for Leibniz parallelisations to a relatively simple algebraic problem -- to find suitable pairs $(E,V)$. One important related question is:\vspace{.2cm}
    
    \emph{What are the possible spaces $M$ that admit a Leibniz parallelisation?}\vspace{.2cm}
    
    \noindent At present, there are not many such spaces known (see the following Section) and it would be interesting to either find new examples or to prove the lack thereof. 
    
    Another interesting situation appears whenever one $E$ admits more than one suitable $V$ -- the corresponding exact elgebroids are then \emph{Poisson--Lie U-dual} \cite{Sakatani, MT}. At present, the only known examples correspond to group manifolds and an important challenge is to find examples beyond this class.
    
\section{Examples}
    We now provide a short list of fairly standard examples, characterised by a pair $(E,V)$.
    \begin{example}
        Taking $E$ to be abelian and $V$ a IIB co-Lagrangian subspace, we get $\ms G_E/\ms G_V\cong T^{n-1}$, the $n-1$-dimensional torus. Other IIB co-Lagrangian subspaces, related to $V$ via $\ms G$-transformations, correspond to Poisson--Lie U-dual setups. This is the standard \emph{U-duality} of toroidal compactifications of the IIB theory.
    \end{example}
    \begin{example}
        Let $\ms H$ be a 1-connected $n-1$-dimensional Lie group, with Lie algebra $\mf h$. We set
        \[E:=\mf h\oplus (S\otimes \mf h^*)\oplus \w{3}\mf h^*\oplus (S\otimes \w{5}\mf h^*),\qquad V:=(S\otimes \mf h^*)\oplus \w{3}\mf h^*\oplus (S\otimes \w{5}\mf h^*),\]
        with the bracket given by the analogue of the formula \eqref{eq:bracket}, i.e.
        \begin{align*}
        [X+\vv\sigma_1+\sigma_3+\vv\sigma_5,&\;X'+\vv\sigma_1'+\sigma_3'+\vv\sigma_5']=\on{ad}_X X'+(\on{ad}_X\vv \sigma_1'-\iota_{X'}\delta\vv\sigma_1)\\
        &+(\on{ad}_X \sigma_3'-\iota_{X'}\delta\sigma_3+\epsilon_{ij}\delta\sigma_1^i\wedge {\sigma'_1}^{\!j})+(\on{ad}_X\vv\sigma_5'-\iota_{X'}\delta\vv\sigma_5+\delta\sigma_3\wedge \vv\sigma_1'-\delta\vv\sigma_1\wedge\sigma_3').\label{eq:bracket}
        \end{align*}
    where $\on{ad}$ is the (co)adjoint action of $\mf h$ and $\delta$ is the Chevalley--Eilenberg differential. We recover $\ms G_E/\ms G_V\cong\ms H$ (or a quotient of $\ms H$ by a discrete subgroup) -- the resulting Leibniz parallelisation \cite{CSCW} is the one induced by the natural trivialisation of $T\ms H$. Obviously, this can be twisted by the algebraic counterparts of $\mc F_1$, $\vv F_{\!3}$, $F_5$. More examples of Leibniz parallelisations over group manifolds were constructed and studied in \cite{Sakatani, MT}.
    \end{example}
    \begin{example}
        Considering the elgebra $E$ from the previous example, we will now show how the generalised Yang--Baxter deformations of \cite{BDMCS,BGM,Sakatani,MT,MST} fit in the present framework. The idea is to deform the subelgebra $V$, while keeping it transverse to $\mf h$ (the resulting quotient space is then again $\ms H$). First, note that any type-IIB co-Lagrangian subspace transverse to $\mf h$ is of the form $g\cdot V$, for some $g\in \ms (S\otimes \w{2}T)\oplus\w{4}T\subset \ms G$. The condition for $g\cdot V$ to define a subelgebra, i.e.
        \[[g\cdot V,g\cdot V]\subset g\cdot V,\]
        is called the \emph{generalised Yang--Baxter equation} ($g$ can be seen as a ``\emph{generalised r-matrix}''). Having a solution to this equation such that the condition $\on{Tr}_E\on{ad}_v=\tfrac{\lambda}{\lambda-1}\on{Tr}_V\on{ad}_v$ holds for the deformed subelgebra, we get a Leibniz parallelisation on $\ms H$. All such parallelisations, including the (trivial) one from the previous example, are by definition related by the Poisson--Lie U-duality.\footnote{The same construction works also in the M-theory case, with the element $g$ belonging to $\w{3}T\oplus \w{6} T$.}
    \end{example}
    \begin{example}\label{ex:sphere}
        Let us now describe an example from \cite{LSCW}.\footnote{To make contact with a more standard notation, we will now relax the condition that $\ms G_E$ is simply-connected.}
        Motivated by the decomposition $\mathbf{27}=(\mathbf{15},\mathbf{1})\oplus (\mathbf{6},\mathbf{2})$ of $E$ under $\mf{sl}(6,\R)\oplus\mf{sl}(2,\R)\subset \mf e_{6(6)}$, we take
        \[E:=\mf{so}(6)\oplus (V_6\oplus V_6),\qquad V:=\mf{so}(5)\oplus (V_6\oplus V_6),\]
        where $V_6$ is the vector representation of $\mf{so}(6)$. The bracket on $E$ is defined as follows: for $u\in\mf{so}(6)$, $[u,\cdot]$ coincides with the $\mf{so}(6)$ representation on $E$, while $[V_6\oplus V_6,E]=0$. 
        Note that this implies $\mc D\neq 0$ and $\ms G_E=\ms{SO}(6)$, $\ms G_V=\ms{SO}(5)$. The resulting space is $\ms{SO}(6)/\ms{SO}(5)\cong S^5$.
    \end{example}
    \begin{example}
        Performing a Wigner--İn\" on\" u contraction on the previous example, we obtain a Leibniz parallelisation from \cite{HS}. Explicitly, this is given by replacing the Lie algebras $\mf{so}(6)$ and $\mf{so}(5)$ by $\mf{so}(5)\ltimes \R^5$ and $\mf{so}(4)\ltimes\R^4$, respectively. We get
        \[E:=(\mf{so}(5)\ltimes \R^5)\oplus (V_6\oplus V_6),\qquad V:=(\mf{so}(4)\ltimes\R^4)\oplus (V_6\oplus V_6),\]
        (Note that $\mf{so}(5)\ltimes \R^5\subset \mf{gl}(6,\R)$, so that $V_6$ still carries an action of $\mf{so}(5)\ltimes \R^5$ and hence the bracket on $E$ is well defined.) The corresponding groups are $\ms G_E=\ms{SO}(5)\ltimes \R^5$, $\ms G_V=\ms{SO}(4)\ltimes \R^4$, yielding $\ms{SO}(5)\ltimes \R^5/\ms{SO}(4)\ltimes \R^4\cong S^4\times \R$.
        
        Keeping the same $E$, we can also use the embedding $\mf{so}(5)\subset \mf{so}(5)\ltimes \R^5$ to choose \[V:=\mf{so}(5)\oplus (V_6\oplus V_6),\] resulting in the quotient space $\R^5$. We thus get a pair of Poisson--Lie U-dual spaces, $S^4\times\R$ and $\R^5$, in analogy to a result \cite{Inverso} in the context of reductions to 4 dimensions.
    \end{example}
\section{Summary and conclusions}
    This note and its companion \cite{one} demonstrate that both M-theory and type-IIB exceptional generalised geometry fit naturally into the framework of elgebroids. This gives a useful tool for investigating various aspects of the related geometry and physics, for instance Leibniz parallelisations and Poisson--Lie U-duality. 
    
    Leibniz parallelisations correspond to maximal consistent truncations -- compactifying 11D or type-IIB supergravity on a Leibniz parallelisable manifold provides a lift for the solutions of the lower-dimensional supergravity to the full 11D/type-IIB theory. However, up to now no classification of Leibniz parallelisable spaces is known. The present framework, together with that of Inverso \cite{Inverso}, gives a direct algebraic method for finding such spaces, that is to address the string-landscape question: \vspace{.2cm} 
    
    \emph{Which embedding tensors (elgebras) can be realised as Leibniz parallelisations?}\vspace{.2cm}
    
    \noindent The answer is that one needs to find an elgebra $E$ together with a co-Lagrangian subelgebra $V\subset E$, which is either
    \begin{itemize}
        \item of codimension $n$, satisfying $\on{Im}\mc D\subset V$,
        \item of codimension $n-1$, satisfying $\on{Im}\mc D\subset V$ and $\on{Tr}_E\on{ad}_v=\tfrac{\lambda}{\lambda-1}\on{Tr}_V\on{ad}_v$ for every $v\in V$.
    \end{itemize}
    This results in a Leibniz parallelisable space in the M-theory and type-IIB case, respectively, given by the quotient of the corresponding groups $\ms G_E/\ms G_V$ (supposing $\ms G_V\subset \ms G_E$ is closed). Fixing $E$ and taking different subelgebras $V$ leads to mutually Poisson--Lie U-dual setups. Notice that if a given elgebra admits co-Lagrangian subelgebras of both dimensions ($n$ and $n-1$), we obtain a duality between IIB and M-theory setups.
    
    Suppose there also exists a Lagrangian subelgebra $W\subset E$, which is transverse to $V$ (and is of complementary dimension). Since $W$ is Lagrangian, its bracket is skew-symmetric and makes $W$ a Lie algebra. In such a case, the quotient $\ms G_E/\ms G_V$ can be identified with a group integrating $W$ (or its discrete quotient). This corresponds to the \emph{exceptional Drinfeld algebra} construction of \cite{Sakatani, MT}.\footnote{The name comes from Poisson--Lie T-duality (without spectators), where instead of the pair $V\subset E$ we take a Lagrangian subalgebra of a quadratic Lie algebra -- having another complementary Lagrangian subalgebra equips the first one with the structure of a Drinfeld double.} Relaxing the assumption of existence of $W$, i.e.\ passing from exceptional Drinfeld algebras to elgebras, allows one to accommodate additional non-group examples, such as the spheres (c.f.\ Example \ref{ex:sphere}).
    
    Note the appearance of the ``trace condition'' in the type-IIB case (with $\on{codim}V=n-1$). This is related to the ``algebraic'' possibility of twisting the bracket of a IIB-exact elgebroid with a pair of vectors $\vv\psi$, the presence of which would however lead to the breakdown of the Jacobi identity. The vanishing of $\vv\psi$ is equivalent to condition \eqref{eq:trace} -- and the latter is in turn equivalent to the above ``trace condition''. Physically, the trace condition is a restriction on the gauging of the ``trombone symmetry'', namely that, using the decomposition of $\mathfrak{g}$ in the Appendix, the embedding tensor is constrained such that, for all $v\in V$, the action of $\on{ad}_v$ can gauge the $\mathfrak{gl}(T)$ factor but not the $\mathbb{R}$ factor. We note that this condition is present also in the M-theory case (with $\on{codim}V=n$), but it is automatically satisfied (i.e.\ it follows from the other constraints).\footnote{This translates to a slight refinement of the conditions in \cite{Inverso}. In particular, our analysis implies that the ``C-constraint'' is identically satisfied for the type M co-Lagrangian subelgebras and is equivalent to the trace condition in the type IIB co-Lagrangian case. We are grateful to Gianluca Inverso for discussions on clarifying this point.} This can be ultimately traced back to the fact that no analogue of the vectors $\vv\psi$ enters in the corresponding analysis.
    
    Finally recall that one can also consider consistent truncations of 11D and IIB supergravity with less or even no supersymmetry. These have a generic description in terms of $\ms G$-structures in generalised geometry \cite{CJPW} (see also \cite{CDFMGI,Malek} for the half-maximal theory) that encode the matter content of the truncated theory. In each case there is an underlying Leibniz sub-algebroid of the M-theory or IIB elgebroid that is defined by a Leibniz algebra, in analogy to the exceptional Leibniz parallelisations discussed here. It would clearly be interesting to investigate how these fit in the general language of $\ms G$-algebroids defined in \cite{one} and if one can find analogues of the structure theorem derived here that would determine which gaugings in theories with less supersymmetry can be realised as consistent truncations. 
    
{\appendix
\section{Details of the algebra}
    Here we follow \cite{AW}.
    First, there is a $\ms{GL}(n-1,\R)\times \ms{SL}(2,\R)$ subgroup of $\ms G:=\ms E_{n(n)}\times\R^+$, under which we have
    \begin{align*}
        \mf{g}&\cong \R\oplus \mf{gl}(T)\oplus \mf{sl}(S)\oplus (S\otimes \w{2}T)\oplus (S\otimes \w{2}T^*)\oplus \w{4}T\oplus\w{4}T^*,\\
        E &\cong T\oplus (S\otimes T^*)\oplus \w{3}T^*\oplus (S\otimes \w{5}T^*),\\
        N &\cong S\oplus \w{2}T^*\oplus (S\otimes \w{4}T^*)\oplus (T^*\otimes\w{5}T^*),
    \end{align*}
    where $T:=\R^{n-1}$ and $S:=\R^2$. The first identification is arranged so that $\mf{gl}(T)$ acts on $\mf g$ in the standard way.\footnote{In particular, the $\mf{e}_{n(n)}$-subalgebra corresponds to taking the value of the $\R$-component equal to $1/(9-n)$ times the trace of the $\mf{gl}(T)$-component.}
    
    Let us now describe the action of $\mf{g}$ on $E$. First, $\mf{gl}(T)$ and $\mf{sl}(S)$ act in the obvious way, while $\R$ acts with weight 1. Writing\footnote{The arrow signifies that the given tensor has value in $S$.} $u=X+\vv \sigma_1+\sigma_3+\vv \sigma_5=X+\sigma$ for an element of $E$, and $\vv w_2+\vv a_2+w_4+a_4\in (S\otimes \w{2}T)\oplus (S\otimes \w{2}T^*)\oplus \w{4}T\oplus\w{4}T^*$, the rest of the action is given by
    \[\vv w_2 \cdot u=\epsilon_{ij}\iota_{\sigma_1^i} w_2^j+\iota_{\vv w_2}\sigma_3+\epsilon_{ij} \iota_{w_2^i}\sigma_5^j,\quad w_4\cdot u=-\iota_{\sigma_3}w_4-\iota_{w_4}\vv\sigma_5,\]
%    \vspace{-0.4cm}
    \[\vv a_2\cdot u=\iota_X\vv a_2+\epsilon_{ij}\sigma_1^i\wedge a_2^j +\sigma_3\wedge\vv a_2,\quad a_4\cdot u=\iota_X a_4-\vv \sigma_1\wedge a_4.\]

    The map $E\otimes E\to N$ is symmetric and is given by
    \[X\otimes \sigma \mapsto \iota_X \sigma,\qquad \vv\sigma_1\otimes \sigma \mapsto \epsilon_{ij} \sigma_1^i \wedge\sigma_1^j-\vv\sigma_1\wedge\sigma_3+\epsilon_{ij}\sigma_1^i\otimes\sigma_5^j,\qquad \sigma_3\otimes\sigma_3\mapsto -\sigma_3\bar \otimes \sigma_3,\]
    where we defined
    \[\bar\otimes\colon \w{3}T^*\otimes \w{3}T^*\to T^*\otimes \w{5}T^*, \qquad \iota_X(\alpha\bar\otimes\beta)=(\iota_X\alpha)\wedge\beta \quad \forall X\in T.\]
    The dual of the second map $N\to E\otimes E$ is given by the exactly analogous formulas, up to an overall factor, which is fixed by the condition \eqref{eq:meta}. Note that $T\subset E$ is Lagrangian and $(S\otimes T^*)\oplus \w{3}T^*\oplus (S\otimes \w{5}T^*)$ is type IIB co-Lagrangian. %Similarly, any pair consisting of a type IIB co-Lagrangian subspace and a complementary Lagrangian subspace can be related to $(V,T)$ via a $\ms G$-transformation (c.f.\ \cite{one}).
    
    Consider now the subalgebra $\mf n\subset \mf{g}$ given by elements which send $E$ into $(S\otimes T^*)\oplus \w{3}T^*\oplus (S\otimes \w{5}T^*)$. One easily sees that 
    \[\mf n=\R'\oplus \mf{sl}(S)\oplus (S\otimes \w{2}T^*)\oplus\w{4}T^*,\qquad \R':=\{(\tfrac{c}{2},-\tfrac{c}{2}\mathds{1})\in \R\oplus \mf{gl}(T)\mid c\in\R\}.\]
    In particular, $\R'$ acts on $T$, $(S\otimes T^*)$, $\w{3}T^*$, and $(S\otimes \w{5}T^*)$ with weights $0$, $1$, $2$, and $3$, respectively.
    For notational convenience we also introduce the subalgebra $\mf{gl}(S)\cong \R'\oplus \mf{sl}(S)\subset \mf g$, with $\R'\ni c\mapsto c \mathds{1}$.
}


\begin{thebibliography}{99}
        \bibitem{AW} A. Ashmore, D. Waldram, \emph{Exceptional Calabi-Yau spaces: the geometry of $\mathcal{N}=2$ backgrounds with flux}, Fortsch. Phys. 65 (2017) 1, 1600109.
        \bibitem{BDMCS} I. Bakhmatov, N. S. Deger, E. T. Musaev, E. Ó Colgáin, M. M. Sheikh-Jabbari, \emph{Tri-vector deformations in $d=11$ supergravity}, JHEP 08 (2019) 126.
        \bibitem{BGM} I. Bakhmatov, K. Gubarev, E. T. Musaev, \emph{Non-abelian tri-vector deformations in $d=11$ supergravity}, JHEP 05 (2020) 113.
        \bibitem{Baraglia} D. Baraglia, \emph{Leibniz algebroids, twistings and exceptional generalized geometry}, J. Geom. Phys. 62 (2012) 903--934.
        \bibitem{BCKT} D. S. Berman, M. Cederwall, A. Kleinschmidt, D. C. Thompson, \emph{The gauge structure of generalised diffeomorphisms}, JHEP 01 (2013) 064.
        \bibitem{BMP} C. D. A. Blair, E. Malek, J. H. Park, \emph{M-theory and Type IIB from a Duality Manifest Action}, JHEP 01 (2014) 172.
        \bibitem{BTZ} C. D. A. Blair, D. C. Thompson, S. Zhidkova, \emph{Exploring Exceptional Drinfeld Geometries}, JHEP 09 (2020) 151.
        \bibitem{BHL} P. du Bosque, F. Hassler, D. Lüst, \emph{Generalized parallelizable spaces from exceptional field theory}, JHEP 01 (2018) 117.
        \bibitem{one} M. Bugden, O. Hulik, F. Valach, D. Waldram, \emph{G-algebroids: a unified framework for exceptional and generalised geometry, and Poisson-Lie duality}, Fortsch. Phys. 69 (2021) 4--5, 2100028.
        \bibitem{CdFPSW} D.~Cassani, O.~de Felice, M.~Petrini, C.~Strickland-Constable and D.~Waldram, \emph{Exceptional Generalised Geometry for Massive IIA and Consistent Reductions}, JHEP 08 (2016) 074.
        \bibitem{CJPW} D.~Cassani, G.~Josse, M.~Petrini and D.~Waldram, \emph{Systematics of Consistent Truncations from Generalised Geometry}, JHEP 11 (2019) 017.
        \bibitem{CDFMGI} F.~Ciceri, G.~Dibitetto, J.~J.~Fernandez-Melgarejo, A.~Guarino and G.~Inverso, \emph{Double Field Theory at $SL(2)$ Angles}, JHEP 05 (2017) 028.
        \bibitem{CGI} F.~Ciceri, A.~Guarino and G.~Inverso, \emph{The Exceptional Story of Massive IIA Supergravity}, JHEP 08 (2016) 154.
        \bibitem{CSCW} A. Coimbra, C. Strickland-Constable, D. Waldram, \emph{$E_{d(d)}\times\mathbb{R}^+$ generalised geometry, connections and M theory}, JHEP 02 (2014) 054.
        \bibitem{CSCW0} A. Coimbra, C. Strickland-Constable, D. Waldram, \emph{Supergravity as Generalised Geometry I: Type II Theories}, JHEP 11 (2011) 091.
        \bibitem{CSCW2} A. Coimbra, C. Strickland-Constable, D. Waldram, \emph{Supergravity as Generalised Geometry II: $E_{d(d)}\times\mathbb{R}^+$ and M theory}, JHEP 03 (2014) 019.
        \bibitem{GMSW} J. P. Gauntlett, D. Martelli, J. Sparks, D. Waldram, \emph{Supersymmetric $AdS_5$ solutions of type IIB supergravity}, Class. Quant. Grav. 23 (2006) 4693--4718.
        \bibitem{HS} O. Hohm, H. Samtleben, \emph{Consistent Kaluza-Klein Truncations via Exceptional Field Theory}, JHEP 01 (2015) 131.
        \bibitem{HS2} O. Hohm, H. Samtleben, \emph{Exceptional Field Theory I: $E_{6(6)}$ covariant Form of M-Theory and Type IIB}, Phys. Rev. D 89 (2014) 6, 066016.
        \bibitem{Hull} C. M. Hull, \emph{Generalised Geometry for M-Theory}, JHEP 07 (2007) 079.
        \bibitem{HZ} C.~Hull and B.~Zwiebach, \emph{The Gauge Algebra of Double Field Theory and Courant Brackets}, JHEP 09 (2009) 090. 
        \bibitem{Inverso} G. Inverso, \emph{Generalised Scherk-Schwarz reductions from gauged supergravity}, JHEP 12 (2017) 124.
        \bibitem{KS} C. Klim\v cík, P. Ševera, \emph{Dual non-Abelian T-duality and the Drinfeld double}. Phys. Lett. B 351 (1995), 455--462.
        \bibitem{LSCW} K. Lee, C. Strickland-Constable, D. Waldram, \emph{Spheres, generalised parallelisability and consistent truncations}, Fortsch. Phys. 65 (2017) 10--11, 1700048.
        \bibitem{LWX} Z. J. Liu, A. Weinstein, P. Xu, \emph{Manin triples for Lie bialgebroids}, J. Diff. Geom. 45 (1997) 3, 547--574.
        \bibitem{Malek} E.~Malek, \emph{Half-Maximal Supersymmetry from Exceptional Field Theory}, Fortsch. Phys. 65 (2017) 10--11, 1700061.
        \bibitem{MST} E. Malek, Y. Sakatani, D. C. Thompson, \emph{$E_{6(6)}$ exceptional Drinfel’d algebras}, JHEP 01 (2021) 020.
        \bibitem{MT} E. Malek, D.C. Thompson, \emph{Poisson-Lie U-duality in exceptional field theory}, JHEP 04 (2020) 058.
        \bibitem{PW} P. P. Pacheco, D. Waldram, \emph{M-theory, exceptional generalised geometry and superpotentials}, JHEP 09 (2008) 123.
        \bibitem{Sakatani} Y. Sakatani, \emph{$U$-duality extension of Drinfel’d double}, Prog Theor Exp Phys (2020), PTEP 2020 (2020) 2, 023B08.
        \bibitem{Sakatani2} Y. Sakatani, \emph{Extended Drinfel’d algebras and non-Abelian duality},  Prog Theor Exp Phys (2021), PTEP 2021 (2021) 6, 063B02.
        \bibitem{Samtleben} H.~Samtleben, \emph{Lectures on Gauged Supergravity and Flux Compactifications}, Class. Quant. Grav. 25 (2008) 214002.
        \bibitem{Strobl} T.~Strobl, \emph{Non-Abelian Gerbes and Enhanced Leibniz Algebras}, Phys. Rev. D 94 (2016) 021702.
        \bibitem{SW} T.~Strobl and F.~Wagemann, \emph{Enhanced Leibniz Algebras: Structure Theorem and Induced Lie 2-Algebra}, Commun. Math. Phys. 376 (2019) 1, 51--79.
        \bibitem{Severa1} P. Ševera, \emph{Letters to Alan Weinstein about Courant algebroids},  arXiv:1707.00265.
        \bibitem{Severa2} P. Ševera, \emph{Poisson-Lie T-Duality and Courant Algebroids}, Lett. Math. Phys. 105 (2015) 12, 1689--1701.
        \bibitem{Severa3} P. Ševera, \emph{Poisson-Lie T-duality as a boundary phenomenon of Chern-Simons theory}, JHEP 05 (2016) 044.
        \bibitem{dWNS} B.~de Wit, H.~Nicolai and H.~Samtleben, \emph{Gauged Supergravities, Tensor Hierarchies, and M-theory}, JHEP 02 (2008) 044.
    \end{thebibliography}
\end{document}